\documentclass{pmet}

\submit

\makeatletter
\@ifundefined{MPFourScale}{\def\MPFourScale{1.00}}{}
\DeclareFontFamily{U}{mp4}{}%
\DeclareFontShape{U}{mp4}{m}{n}{<->s * [\MPFourScale]cmb10}{}
\DeclareSymbolFont{boldgreekuc}{U}{mp4}{m}{n}
\DeclareMathSymbol{\bfAlpha}{\mathord}{boldgreekuc}{"41}
\DeclareMathSymbol{\bfBeta}{\mathord}{boldgreekuc}{"42}
\DeclareMathSymbol{\bfPsi}{\mathord}{boldgreekuc}{"09}
\DeclareMathSymbol{\bfDelta}{\mathord}{boldgreekuc}{"01}
\DeclareMathSymbol{\bfEpsilon}{\mathord}{boldgreekuc}{"45}
\DeclareMathSymbol{\bfPhi}{\mathord}{boldgreekuc}{"08}
\DeclareMathSymbol{\bfGamma}{\mathord}{boldgreekuc}{"00}
\DeclareMathSymbol{\bfEta}{\mathord}{boldgreekuc}{"48}
\DeclareMathSymbol{\bfIota}{\mathord}{boldgreekuc}{"49}
\DeclareMathSymbol{\bfXi}{\mathord}{boldgreekuc}{"04}
\DeclareMathSymbol{\bfKappa}{\mathord}{boldgreekuc}{"4B}
\DeclareMathSymbol{\bfLambda}{\mathord}{boldgreekuc}{"03}
\DeclareMathSymbol{\bfMu}{\mathord}{boldgreekuc}{"4D}
\DeclareMathSymbol{\bfNu}{\mathord}{boldgreekuc}{"4E}
\DeclareMathSymbol{\bfPi}{\mathord}{boldgreekuc}{"05}
\DeclareMathSymbol{\bfTheta}{\mathord}{boldgreekuc}{"02}
\DeclareMathSymbol{\bfRho}{\mathord}{boldgreekuc}{"52}
\DeclareMathSymbol{\bfSigma}{\mathord}{boldgreekuc}{"06}
\DeclareMathSymbol{\bfTau}{\mathord}{boldgreekuc}{"54}
\DeclareMathSymbol{\bfVartheta}{\mathord}{boldgreekuc}{"02} 
\DeclareMathSymbol{\bfOmega}{\mathord}{boldgreekuc}{"0A}
\DeclareMathSymbol{\bfVarphi}{\mathord}{boldgreekuc}{"08} 
\DeclareMathSymbol{\bfUpsilon}{\mathord}{boldgreekuc}{"07}
\DeclareMathSymbol{\bfZeta}{\mathord}{boldgreekuc}{"5A}

\DeclareFontFamily{U}{mp4sl}{}%
\DeclareFontShape{U}{mp4sl}{m}{n}{<->s * [\MPFourScale]cmmib10}{}
\DeclareSymbolFont{boldgreek}{U}{mp4sl}{m}{n}
\DeclareMathSymbol{\bfalpha}{\mathord}{boldgreek}{"0B}
\DeclareMathSymbol{\bfbeta}{\mathord}{boldgreek}{"0C}
\DeclareMathSymbol{\bfpsi}{\mathord}{boldgreek}{"20}
\DeclareMathSymbol{\bfdelta}{\mathord}{boldgreek}{"0E}
\DeclareMathSymbol{\bfepsilon}{\mathord}{boldgreek}{"0F}
\DeclareMathSymbol{\bfphi}{\mathord}{boldgreek}{"1E}
\DeclareMathSymbol{\bfgamma}{\mathord}{boldgreek}{"0D}
\DeclareMathSymbol{\bfeta}{\mathord}{boldgreek}{"11}
\DeclareMathSymbol{\bfiota}{\mathord}{boldgreek}{"13}
\DeclareMathSymbol{\bfxi}{\mathord}{boldgreek}{"18}
\DeclareMathSymbol{\bfkappa}{\mathord}{boldgreek}{"14}
\DeclareMathSymbol{\bflambda}{\mathord}{boldgreek}{"15}
\DeclareMathSymbol{\bfmu}{\mathord}{boldgreek}{"16}
\DeclareMathSymbol{\bfnu}{\mathord}{boldgreek}{"17}
\DeclareMathSymbol{\bfpi}{\mathord}{boldgreek}{"19}
\DeclareMathSymbol{\bfvartheta}{\mathord}{boldgreek}{"23}
\DeclareMathSymbol{\bfrho}{\mathord}{boldgreek}{"1A}
\DeclareMathSymbol{\bfsigma}{\mathord}{boldgreek}{"1B}
\DeclareMathSymbol{\bftau}{\mathord}{boldgreek}{"1C}
\DeclareMathSymbol{\bftheta}{\mathord}{boldgreek}{"12}
\DeclareMathSymbol{\bfomega}{\mathord}{boldgreek}{"21}
\DeclareMathSymbol{\bfvarphi}{\mathord}{boldgreek}{"27}
\DeclareMathSymbol{\bfchi}{\mathord}{boldgreek}{"1F}
\DeclareMathSymbol{\bfupsilon}{\mathord}{boldgreek}{"1D}
\DeclareMathSymbol{\bfzeta}{\mathord}{boldgreek}{"10}

\makeatother

\usepackage[dvips]{graphicx}
\usepackage{amsmath}
\usepackage{arydshln}
\usepackage{algorithm}
\usepackage{algorithmic}
\usepackage{booktabs}
\usepackage{multirow}
\usepackage{subfigure}

\newtheorem{theorem}{Theorem}
\renewcommand{\normalsize}{\fontsize{11}{16}\selectfont}
\renewcommand{\footnotesize}{\fontsize{10}{12}\selectfont}
\fontsize{11}{16}\selectfont

\begin{document}

\title{Two-Step Iterative GMM Structure for Estimating Mixed Correlation Coefficient Matrix}

\maketitle


\author{Liu Ben}

\affil{Zhejiang University}



\reprints{Correspondence should be sent to\\

\noindent E-Mail: 12035024@zju.edu.cn \\
\noindent Phone: +86 17816878025 \\
\noindent Fax:  \\
\noindent Website:  \\}

\newpage\vspace*{24pt}

\RepeatTitle{Two-Step Iterative GMM Structure for Estimating Mixed Correlation Coefficient Matrix}

\begin{center}\vskip3pt

\vspace{32pt}

Abstract\vskip3pt

\end{center}

\begin{abstract}

In this article, we propose a new method for calculating the mixed correlation coefficient (Pearson, polyserial and polychoric) matrix and its covariance matrix based on the GMM framework. We build moment equations for each coefficient and align them together, then solve the system with Two-Step IGMM algorithm. Theory and simulation show that this estimation has consistency and asymptotic normality, and its efficiency is asymptotically equivalent to MLE. Moreover, it is much faster and the model setting is more flexible (the equations for each coefficient are blocked designed, you can only include the coefficients of interest instead of the entire correlation matrix), which can be a better initial estimation for structural equation model.

\end{abstract}

\vspace{\fill}\newpage

\section{Introduction}
In researches in behavioral sciences, education, psychology, medicine, etc., it is common that the observed variables are measured on ordinal scales. For example, in Likert, Roslow, and Murphy (1934), Likert scale is proposed to assess responses in surveys, allowing individuals to express how much respondents agree or disagree with a particular statement on a five (or seven) point scale. In Hartrick, Kovan, Shapiro (2003), Numeric Rating Scale  (NRS-11) uses an 11-point scale to assess patient self-reporting of pain, with 0 no pain, 1-3 mild pain (sleep is not affected), 4-6 moderate pain (sleep is affected) and 7-10 severe pain (sleep is seriously affected).  These ordinal variables can be viewed as linked to hypothetical, latent continuous variables such as the degree of agreement or pain. In the social sciences, there are also many real, observable continuous variables that are deliberately coarsened for some purpose. For instance, when asking questions about sensitive or personal quantitative attributes (income, alcohol consumption, age), to reduce the non-response rate, researchers may only ask the respondent to select one of two very broad categories (under or over \$30K, 65 years old, etc.). Pointed by Robitzsch (2020), one can analyze ordinal data by assigning integer values to each category, then proceed in the analysis as if the data had been measured on an interval scale with desired distributional properties.  

Another popular approach is to parameterize underlying continuous variables and build latent variable models, see Moustaki (2000), which can model a small number of latent variables to describe the observed ordinal variables. The most common choice for the latent distribution is normal whose information can be captured by the first two moments. After standardizing the latent variables, the parameters to be estimated are only the correlation coefficient matrix, which can be estimated using a bivariate normal distribution separately. Supposing $Y_1, Y_2$ two continuous variables, $X_1, X_2$ two ordinal variables related to continuous latent $Z_1, Z_2$, and $Y_1, Y_2, Z_1, Z_2$ follows the multivariate normal distribution, then correlation between $Y$s is the well-known Pearson correlation coefficient. Suggested by Pearson (1900), the correlation between $Y$ and $Z$ is called polyserial correlation coefficient; when $X$s all have only two categories, the correlation between $X$s is called tetrachoric correlation based on 2×2 contingency table, which was generalized to the polychoric correlation for the case where $X$s have multiple categories by Ritchie-Scott (1918) and Pearson and Pearson (1922) in the early 20th century.

There are several obvious shortcomings in using Pearson correlation coefficient to analyze ordinal variables. The only information provided by this ordinal scale variables is the number of subjects in each categories(cells) in a contingency table, in this case, Pearson correlation coefficient cannot even be calculated; if the Pearson correlation is used, the relationship between measures would be artificially restricted due to the restrictions imposed by categorization (Gilley and Uhlig (1993)), just like Spearman or Kendall correlation coefficients, the value of the latent continuous variable does not affect the observed ordinal score as long as the relative order is the same.

A widespread application of correlation coefficient matrices is factor analysis. In exploratory factor analysis (EFA) and confirmatory factor analysis (CFA), Holgado-Tello et al. (2010) discusses the superiority of polychoric correlation. Extended factor analysis, the polychoric correlation matrix is very important for parameter estimation and inference of structural equation models with ordinal responses. In Muthén (1984) and Jöreskog (1994), the latent variables in SEM were assumed to be normal, and wrote the correlation matrix as a function of parameters through the measurement and structural model, to minimize the distance between it and the sample correlation matrix. Thus, the estimator of SEM is actually a function of the sample correlation matrix. and its asymptotic distribution depends on the asymptotic mean and covariance of the mixed correlation matrix (including Pearson, polyserial, polychoric correlation).

More recently, psychological science has been combined with networks model, which is called psychological networks by researchers. In Epskamp, Borsboom, and Fried (2018), psychological networks has been used in various different fields of psychology. Lauritzen (1996) proposed the Gaussian graphical model (GGM), in which edges were directly interpreted as partial correlation coefficient. Thus GGM requires covariance matrix as input for parameter estimation, and polyserial and polychoric correlations can also be used in case the data are ordinal.

Therefore, it is necessary to estimate the value of the mixed correlated matrix and its asymptotic distribution. As previous notation, the product moment correlation between $X$s and $Y$s is called the point polyserial correlation, while the correlation between $Z$s and $Y$s is called the polyserial correlation. Cox (1974) proposed the maximum likelihood estimate (MLE) of the polyserial correlation. Olsson (1979) presented the so-called two-step method to estimating the MLE of the polychoric correlation: first estimates the unknown thresholds of $X$s from the marginal frequencies of the table and then finds the MLE of $\rho$s conditioning on the estimated thresholds. Olsson, Drasgow, and Dorans (1982) derived the relationship between the polyserial and point polyserial correlation and compared the MLE of polyserial correlation with a two-step estimator and with a computationally convenient ad hoc estimator. Generalizing the bivariate calculation, Poon and Lee (1987) proposed a improved MLE-based algorithm to enable simultaneous inference of multi-variable correlation matrices.

Above MLE based methods has been implemented in many popular statistical softwares such as Mplus by Muthén and Muthén (2005), LISREL Jöreskog (2005), PROC CORR in SAS (with POLYCHORIC or OUTPLC = options) see SAS Institute (2017), polycor package in R by Fox (2010) and psych package in R by Revelle (2017), POLYCORR program by Uebersax (2010),and an extensive list of software for computing the polychoric correlation by Uebersax (2006). 

Bayesian method can be regarded as an improvement of MLE. Albert (1992) estimate polychoric correlation coefficient from the Bayesian point of view by using the Gibbs sampler. One attractive feature of this method is that it can be generalized in a straight forward manner to handle a number of nonnormal latent distributions, like bivariate lognormal and bivariate $t$ latent distributions in their simulations. Chen and Choi (2009) and Choi et al. (2011) proposed two new Bayesian estimators, maximum a posteriori (MAP) and expected a posteriori (EAP), and compared them to the ML method.   

The aforementioned methods are all based on likelihood function which involves computations of the double or even multiple normal distribution integrals numerically and requires a very large amount of computational resources. Lee and Poon (1987) considered a GMM structure, but it was only applied on polychoric situation and still not avoided the multiple integrals. Zhang, Liu and Pan (2023) proposed a iteratively reweighted 
least square method to estimate the bivariate polyserial and polychoric correlation. It converted the bivariate integrals into margin single integrals by conditional expectation, thus is computationally fast, but it has not been extended to the asymptotic distribution estimation of multi-variable mixed correlation coefficient matrices. 

In this article, we propose a Generalized Method of Moments (Hansen (1982)) framework to include all three types of correlation coefficients (Pearson, polyserial, polychoric) and design both simultaneous and two-step algorithm to get the consistent estimator and asymptotical distribution of the mixed correlation matrix. We use Legendre polynomials to approximate the normal integral to speed up the calculation (Drezner and Wesolowsky (1990)). We put the algorithm details in section 2, simulation in section 3 and real data analysis in section 4 to compare the proposed algorithm with the ML method in accuray and speed. At last, we conclude with discussions and some works can be done in the future to improve the proposed method.

\section{Methodology}

In this section, we first parameterize both continuous and ordinal variables; then write the random variables and parameters in generalized moment equations; give a one-step method to estimate all parameters simultaneously, as well as a two-step method to estimate nuisance thresholds first, then estimate the correlation coefficient; Finally, we discuss the statistical properties of the estimator. Appendix1 gives a simple example to facilitate understanding of each formula.

\subsection{The Mixed Correlation Coefficient Matrix}

For convenience of notation, we assume that normal variables are all standardized. We model the ordinal variable with Pearson's theory. Let $Y_i$s denote the observed normal variables, $X_i$s denote the observed ordinal variables, $Z_i$s are the corresponding latent normal variable, that's:
\begin{equation}
    X_i = j \text{ if } a_{i,j-1}<Z_i\leq a_{i,j} \text{ for }  j=1,...,s_i
\end{equation}
with $s_i$ the number of categories of $X_i$, $a_{i,j}$s the corresponding thresholds of each category. Specifically, $a_{i,0}=-\infty, a_{i,s_i}=\infty$. 

$Y_i$s,$X_i$s are observed while $Z_i$s not. Then the correlation coefficients between $Y_i$s are the usual Pearson coefficient, between latent $Z_i$s and $Y_i$s are the so-called polyserial coefficients, among latent $Z_i$s are polychoric coefficients. Suppose there are $c$ continuous and $d$ discrete variables, align them as random vector $\mathrm{Y},\mathrm{X},\mathrm{Z}$, then there is a mixed correlation matrix $\mathbf{R}_{(c+d)\times (c+d)}$ containing three types coefficients:
\begin{equation}
    \begin{pmatrix}
        \mathrm{Y}\\
        \mathrm{X}
    \end{pmatrix}\Rightarrow
    \begin{pmatrix}
        \mathrm{Y}\\
        \mathrm{Z}
    \end{pmatrix}\sim
    N(0,\mathbf{R}), \mathbf{R} = 
    \begin{bmatrix}
        \mathbf{R}_{Y}\\
        \mathbf{R}_{XY} & \mathbf{R}_{X}
    \end{bmatrix}
\end{equation}
with ${\mathbf{R}_{Y}}_{c\times c}$ the Pearson matrix, ${\mathbf{R}_{XY}}_{d\times c}$ the polyserial matrix, ${\mathbf{R}_{X}}_{d\times d}$ the polychoric matrix. Since $\mathbf{R}$ is symmetric positive semi-definite with diagonal equal to 1, there are $c(c-1)/2$ Pearson, $cd$ polyserial, $d(d-1)/2$ polychoric coefficients. The total number of parameters of interest to estimate and inference is $(c+d)(c+d-1)/2$. In addition, the thresholds $a_{i,j}$s for segmenting the latent $\mathrm{Z}$ into $\mathrm {X}$ are nuisance parameters, with a number of $\sum_{i=1}^{d}(s_i-1)$. Flatten the lower triangular $\mathbf{R}_Y$, $\mathbf{R}_{XY}$, and lower triangular $\mathbf{R}_X$ by columns into vector $R$, the purpose is to get $\hat{R}$ and its asymptotic distribution.

\subsection{GMM Equation System}
For convenience, we denote the indicator variables as:
\begin{equation*}
\begin{aligned}
    \mathrm{I}(X_j=k)&:=\mathrm{I}_k(X_j)\\
    \mathrm{I}(X_i=k,X_j=l)&:=\mathrm{I}_{kl}(X_iX_j),
\end{aligned}
\end{equation*}
then add the subscripts $yy,yx,xx$ to $\rho$ to represent the coefficients of Pearson, polyserial, and polychoric respectively.

For Pearson coefficient, an equation can be built by directly cross moments:
\begin{equation*}
    \mathrm{E}[Y_iY_j]=\rho^{yy}_{ij}.
\end{equation*}
For polyserial, we calculate the cross moments of $Y_i$s and the indicator of $X_i$s:
\begin{equation*}
    \mathrm{E}[Y_i\mathrm{I}_k(X_j)]=\rho_{ij}^{yx} \mathrm{E}[Z_j\mathrm{I}_k(X_j)]=\rho_{ij}^{yx} \int_{a_{j,k-1}}^{a_{j,k}}z\phi(z)\mathrm{d} z = \rho_{ij}^{yx}(\phi(a_{j,k-1})-\phi(a_{j,k})) :=\rho_{ij}^{yx} \xi_{j,k},
\end{equation*}
with $\phi(z)$ the standard normal density. For polychoric, the cross moments of indicators is also the probability of binary normal variables fall in thresholds rectangles:
\begin{equation*}
\begin{aligned}
    \mathrm{E}[\mathrm{I}_{kl}(X_i,X_j)]&=P(a_{i,k-1}<Z_i\leq a_{i,k},a_{j,l-1}<Z_j\leq a_{j,l})\\
    &=\int_{a_{i,k-1}}^{a_{i,k}}\int_{a_{j,l-1}}^{a_{j,l}}\phi(z_1,z_2;\rho_{ij}^{xx})\mathrm{d}z_1\mathrm{d}z_2:=\Phi_{kl}(\rho_{ij}^{xx})
\end{aligned}
\end{equation*}
with $\phi(z_1,z_2)$ the standard bivariate normal density with a correlation coefficient $\rho^{xx}_{ij}$. This setting is similar to that in Lee and Poon (1987), but they used the cell probability of a multiple contingency table while this article only uses bivariates. The advantage is that the modeling is more flexible, and you don’t need to include the corresponding equations for some $\rho_{i_2 j_2}^{xx}$ that you don’t care about. In addition, the double integrals can be accelerated very well. Now align all the expectation into a generalized moment equations system:
\begin{equation}
    \mathrm{E}
\begin{bmatrix}
    Y_{i_1}Y_{j_1}-\rho_{i_1j_1}^{yy}\\
    Y_{i_1}\mathrm{I}_{k}(X_{i_2})-\rho_{i_1 i_2}^{yx}\xi_{i_2,k}\\
    \mathrm{I}_{kl}(X_{i_2}X_{j_2})-\Phi_{kl}(\rho_{i_2 j_2}^{xx})
\end{bmatrix}
:=\mathrm{E}[g(R,a)]
=0,
\end{equation}
with subscripts:
\begin{equation*}
\begin{cases}
1 \leq i_1 \leq c, 1\leq j_1<i_1;\\
1 \leq i_2 \leq d, 1 \leq k \leq s_{i_2};\\
1 \leq j_2 < i_2, 1 \leq l \leq s_{j_2}.
\end{cases} 
\end{equation*}

Notice that the thresholds in $g(\cdot)$ are also unknown, we write them in equations:
\begin{equation}\label{momenth}
    \mathrm{E}
\begin{bmatrix}
    \mathrm{I}_{k}(X_{i_2})-(\Phi(a_{i_2,k})-\Phi(a_{i_2,k-1}))
\end{bmatrix}:= \mathrm{E}[h(a)]=0,
\end{equation}
with $\Phi$ the standard normal distribution function. Putting both $g(\cdot)$ and $h(\cdot)$ together, we get the total population parameters moment equations:
\begin{equation}\label{totalmoment}
\mathrm{E}[u(a,R)]=
\mathrm{E}
    \begin{bmatrix}
        h(a)\\
        g(R,a)
    \end{bmatrix}
    =0.
\end{equation}

Considering the i.i.d. samples $\mathbf{X}_i =(Y_{1i},...,Y_{ci},X_{1i},...,X_{di})',i=1,...,n$, recording $(a,R)$ as $\theta$, by replacing population moments\ref{totalmoment} with sample mean, we got:
\begin{equation}\label{samplemoment}
    m(\mathbf{X};\theta) = \mathrm{E}_n[u] = \frac{1}{n}\sum_{i=1}^{n}
[u(\mathbf{X}_i;\theta)]
\end{equation}
Since $m(\cdot)$ is over-determined most of the time thus cannot be solved to 0, so we minimize the norm with respect to a weight matrix $W$:
\begin{equation}\label{loss}
    \hat{\theta}=\text{arg min} \frac{1}{2}||m(\mathbf{X};\theta)||_W^2= \text{arg min}\frac{1}{2}m'Wm = \text{arg min} L(\theta)
\end{equation}
while $W$ is symmetric positive semi-definite matrix. To get the most efficient estimator, we set:
\begin{equation}\label{Wopt}
    W_{opt} = [\text{Var } u(\cdot)]^{-1}=\mathrm{E}^{-1}[uu']:=\Omega^{-1}. 
\end{equation}
But there are some singularity issues when using the variance of $u$ as above optimal $W$. For the equations in $h(\cdot)$ of each $X_{i_2}$:
\begin{equation*}
    \sum_{k=1}^{s_{i_2}}[\mathrm{I}_{k}(X_{i_2})-\Phi(a_{i_2,k})+\Phi(a_{i_2,k-1})]=0,
\end{equation*}
for polyserial equations in $g(\cdot)$ related to a $Y_{i_1}$:
\begin{equation*}
    \sum_{k=1}^{s_{i_2}}
    [Y_{i_1}\mathrm{I}_{k}(X_{i_2})-\rho_{i_1 i_2}^{yx}\xi_{i_2,k}]
    =Y_{i_1},
\end{equation*}
for polychoric equations in $g(\cdot)$ related to a pair of $X_{i_2},X_{j_2}$:
\begin{equation*}
    \sum_{k=1}^{s_{i_2}}\sum_{l=1}^{s_{j_2}}=
    [\mathrm{I}_{kl}(X_{i_2}X_{j_2})-\Phi_{kl}(\rho_{i_2 j_2}^{yx})]
    =0.
\end{equation*}
Thus, three parts of $u(\cdot)$ have the collinearity problem. A convenient strategy is to replace the inverse matrix with a generalized inverse, so that all conclusions are applicable in parameter estimation and inference, but this will consume more calculations. The suitable solution is to remove redundant equations, that's, each $X_{i_2}$'s thresholds, polyserial pair, polychoric pair, remove one equation (leaving a complete set for each $Y_{i_1}$ in polyserial pair).

Notice that the equations system\ref{totalmoment} is blocked according to $\rho$s thus can be designed freely, while most MLEs will estimate $R$ simultaneously. Counting each block, threshold block for $X_{i_2}$ has $s_{i_2}-1$, Pearson block for $\rho_{i_1 j_1}^{yy}$ has $1$, polyserial block for $\rho_{i_1 i_2}^{yx}$ has $s_{i_2}-1$, polychoric block for $\rho_{i_2 j_2}^{xx}$ has $s_{i_2} s_{j_2} -1$ equations. The total equations in system\ref{totalmoment} is:
\begin{equation*}
    \sum_{i_2 =1}^d(s_{i_2}-1) + \frac{1}{2}c(c-1)+ c[\sum_{i_2=1}^d(s_{i_2}-1)+1] + \sum_{i_2=1}^d\sum_{j_2=2}^{i_2 -1}(s_{i_2}s_{j_2}-1).
\end{equation*}
This is the maximum equations set, the blocks for polyserial and polychoric are over-determined. Correspondingly, there is a minimum equation set, that is, only one equation is retained for each $\rho$. The inspiration is that when original information is missing (for example, only the categorical means are given, the contingency table is incomplete, etc.), we can still incorporate the remaining information into the equation system and obtain a relatively inefficient but consistent estimator, as long as the system is larger than minimum set.

\subsection{Parameters Estimation}
\subsubsection{One-Step Estimator}
Since $\Omega$ is the covariance matrix of $u(\cdot)$, analytically it is also a function of parameters $(R,a)$. An idea is to regard it as a part to be optimized and solve $L(\theta)$ at the same time. However, writing the specific formula of $\Omega$ is very complicated and involves up to quadruple  integrals, so we estimate it with the sample covariance matrix, treat it as a constant, and use iterative GMM to adjust it. That's, for a fixed $(\hat{R},\hat{a})$, set:
\begin{equation}\label{weightmatrix}
    \hat{\Omega} = \mathrm{E}_n[uu'] = \frac{1}{n}\sum_{i=1}^n u(\mathbf{X}_i;\hat{R},\hat{a})u'(\mathbf{X}_i;\hat{R},\hat{a})=\hat{W}^{-1}_{opt},
\end{equation}
minimize $L(\theta)$ with certain method to get new estimator, and repeat the process. Since the loss has good convexity, the complex calculation of Hessian matrix can be dropped and optimization related to gradient (such as gradient descent, BFGS, etc.) can be used. The gradient is:
\begin{equation}\label{gradient}
    \frac{\partial L}{\partial \theta} = G'Wm,
\end{equation}
with $G$ the derivative of $m$. The gradient can be segmented as: 
\begin{equation*}
    G = \frac{\partial m}{\partial \theta}=\mathrm{E}_n[\frac{\partial u}{\partial \theta}]=
    \mathrm{E}_n
    \begin{bmatrix}
        {\partial h}/{\partial a} & {\partial h}/{\partial R}\\
        {\partial g}/{\partial a} & {\partial g}/{\partial R}
    \end{bmatrix}:=
    \begin{bmatrix}
        G_{11} & G_{12}\\
        G_{21} & G_{22}
    \end{bmatrix}.
\end{equation*}
We put the detailed formula for $G$ in Appendix1\ref{app1} and a clear example in Appendix2\ref{app2}. A simultaneous one-step algorithm is designed as Algorithm\ref{alg1}1.
\begin{algorithm}\label{alg1}
\renewcommand{\algorithmicrequire}{\textbf{Input:}}
\renewcommand{\algorithmicensure}{\textbf{Output:}}
\caption{One-Step Iterative GMM}
\begin{algorithmic}[1] 
\REQUIRE $i.i.d.$ samples $\mathbf{X}_i =(Y_{1i},...,Y_{ci},X_{1i},...,X_{di})',i=1,...,n$
\ENSURE $\hat{\theta}:=(\hat{a},\hat{R})$, Var($\hat{\theta}$)
\STATE Initialized $\hat{a}^{(0)}$ as formula (\ref{thresholdsolve}), $\hat{R}^{(0)}$ with Pearson correlation, $\hat{W}^{(0)} = \mathbf{I}$
\STATE Initialized $t = 1$, maxiter = 100, diff = 1, $\epsilon$ = 1e-8 
\WHILE{$t < \text{maxiter} ~\&~ \text{diff} > \epsilon$}
\STATE Solve $\hat{\theta}^{(t)} = \text{arg min}L(\mathbf{X},\hat{W}^{(t-1)};\theta)$ by gradient descent method with initial value $\hat{\theta}^{(t-1)}$. The moments equations can be calculates as formula (\ref{samplemoment}), the gradient as formula (\ref{gradient}) and Appendix1\ref{app1}
\STATE Calculate $\hat{W}^{(t)}$ as formula (\ref{weightmatrix})
\STATE Compute $\text{diff}=||\hat{\theta}^{(t)} - \hat{\theta}^{(t-1)}||_2$. Update $t=t+1$
\ENDWHILE
\STATE Set $\hat{\theta}=\hat{\theta}^{(t)},\hat{W}=\hat{W}^{(t)}$. Calculate $G$ as formula (\ref{gradient}) 
\STATE Calculate the asymptotic covariance matrix as formula (\ref{1stepvar})
\RETURN $\hat{\theta}$, Var($\hat{\theta}$)
\end{algorithmic}
\end{algorithm}

\subsubsection{Two-Step Estimator}
Since we only focus on the properties of $R$, for $a$, we can conduct a simple estimator first, then treat $a$ as constant in $g(\cdot)$ to reduce the number of equations and simplify the estimation. Notice that moment equations (\ref{momenth}) are just-determined, noted $\mathrm{E}_n[\mathrm{I}_{k}(X_{i_2})]=p_{k,i_2}$, the thresholds can be solved as:
\begin{equation}\label{thresholdsolve}
\mathrm{E}_n[h(a)]=0 \Rightarrow
\hat{a}_{i_2,k} = \Phi^{-1}(\sum_{j=1}^{k} p_{{i_2},j}), k=1,..,s_{i_2}-1.
\end{equation}

Then the GMM framework can be rewritten as:
\begin{equation}\label{2stepmoment}
    \begin{gathered}
        \mathrm{E}[g(R)]=0\\
        m(\mathbf{X},\hat{a};R)=\mathrm{E}_n[g]=\frac{1}{n}\sum_{i=1}^{n}
[g(\mathbf{X}_i,\hat{a};R)]\\
\hat{R} = \text{arg min} \frac{1}{2}m'Wm=\text{arg min} L(R)\\
W_{opt} = [\text{Var } g(\cdot)]^{-1}=\mathrm{E}^{-1}[gg']:=\Omega^{-1}.
    \end{gathered}
\end{equation}
And the gradient matrix required in optimization also becomes $G_{22}$. We recommend the two-step method since it reduces the calculation amount of operations such as matrix inversion, but its asymptotic distribution needs to be modified considering the variability of $\hat{a}$. The two-step algorithm is shown as Algorithm 2\ref{alg2}.

\begin{algorithm}\label{alg2}
\renewcommand{\algorithmicrequire}{\textbf{Input:}}
\renewcommand{\algorithmicensure}{\textbf{Output:}}
\caption{Two-Step Iterative GMM}
\begin{algorithmic}[1] 
\REQUIRE $i.i.d.$ samples $\mathbf{X}_i =(Y_{1i},...,Y_{ci},X_{1i},...,X_{di})',i=1,...,n$
\ENSURE $\hat{a},\hat{R}$, Var($\hat{R}$)
\STATE Calculate $\hat{a}$ as formula (\ref{thresholdsolve}), $\hat{\Sigma}$ as formula (\ref{Sigma}). Initialize $\hat{R}^{(0)}$ with Pearson correlation, $\hat{W}^{(0)} = \mathbf{I}$
\STATE Initialized $t = 1$, maxiter = 100, diff = 1, $\epsilon$ = 1e-8 
\WHILE{$t < \text{maxiter} ~\&~ \text{diff} > \epsilon$}
\STATE Slove $\hat{R}^{(t)} = \text{arg min}L(\mathbf{X},\hat{a},\hat{W}^{(t-1)};R)$ by gradient descent method with initial value $\hat{R}^{(t-1)}$. The moments equations can be calculates as formula (\ref{2stepmoment}), the gradient $G_{22}$ as formula (\ref{gradient}) and Appendix1\ref{app1}
\STATE Calculate $\hat{W}^{(t)}$ as formula (\ref{2stepmoment})
\STATE Compute $\text{diff}=||\hat{R}^{(t)} - \hat{R}^{(t-1)}||_2$. Update $t=t+1$
\ENDWHILE
\STATE Set $\hat{R}=\hat{R}^{(t)},\hat{W}=\hat{W}^{(t)}$. Calculate $G_{21},G_{22}$ as Appendix1\ref{app1}
\STATE Calculate asymptotic covariance matrix of $\hat{R}$ as formula (\ref{2stepvar})
\RETURN $\hat{a}$, $\hat{R}$, Var($\hat{R}$)
\end{algorithmic}
\end{algorithm}

\subsubsection{Double Integral Acceleration}
An important reason for restricting moment equations to pairwise relationships between variables is that only double integrals are involved in the calculation. According to Drezner and Wesolowsky (1990), the standard bivariate normal integral can be approximated with Legendre polynomials:
\begin{equation*}
    P(X\leq x,Y\leq y)= \Phi(x,y;\rho) \hat{=} 
    \frac{1}{2}\rho [\phi(x,y;\frac{3-\sqrt{3}}{6}\rho)+\phi(x,y;\frac{3+\sqrt{3}}{6}\rho)]+\Phi(x)\Phi(y).
\end{equation*}
In second order or in more accurate third order:
\begin{equation*}
    \Phi(x,y;\rho) \hat{=} 
    \frac{1}{18}\rho [5\phi(x,y;\frac{1-\sqrt{{3}/{5}}}{2}\rho)+8\phi(x,y;\frac{1}{2}\rho)+5\phi(x,y;\frac{1+\sqrt{{3}/{5}}}{2}\rho)]+\Phi(x)\Phi(y).
\end{equation*}
This can avoid the double integral of complex functions while maintaining a very high accuracy, greatly speeding up the entire parameter estimation process. 

\subsection{Asymptotic Properties}
In this section, we give the large sample properties of the aforementioned one-step and two-step estimators. Most of the conclusions are direct applications of the theory in Newey and McFadden (1994), so we explain them briefly. Denote the true value of parameters are $(a_0,R_0)$.
\begin{theorem}
When $n\rightarrow \infty$, $G$ defined as Appendix1\ref{app1}, $W$ defined as (\ref{Wopt}), for one-step estimator in Algorithm 1\ref{alg1}:
\begin{equation}\label{1stepvar}
    \hat{\theta}\xrightarrow{p} \theta _0
    \text{ and }
    \sqrt{n}(\hat{\theta} - \theta _0) \xrightarrow{d} N(0,(G'WG)^{-1}).
\end{equation}
\end{theorem}
Since one-step framework satisfies the sufficient conditions for consistency and asymptotic normality of GMM:
\begin{itemize}
    \item Moment equation (\ref{totalmoment}) for each parameters is either just-determined or over-determined;
    \item $\hat{\Omega}^{-1}$ in (\ref{weightmatrix}) is consistent estimator for positive definite matrix $W$;
    \item The space of possible $\theta$ is compact;
    \item $u(\cdot)$ is continuous at each $\theta$ with probability one;
    \item Since all integral are bounded by normal density, $\mathrm{E}[||u(\theta)||],\mathrm{E}[||u(\theta)||^2],\mathrm{E}[||G(\theta)||]<\infty$;
    \item $G'WG$ is nonsingular.    
\end{itemize}
And the estimated values for $G,W$ can be obtained by substituting $\hat{\theta}$ into corresponding formula.
\begin{theorem}\label{the2}
When $n\rightarrow \infty$, gradients defined as Appendix1\ref{app1}, $W$ defined as (\ref{2stepmoment}). Denote $\Lambda=(G_{22}'WG_{22})^{-1},\Gamma = G_{22}'WG_{21},\Sigma = \text{Var }h(\cdot) = \mathrm{E}[hh']$, then: for two-step estimator in Algorithm 2\ref{alg2}:
\begin{equation}\label{2stepvar}
\begin{aligned}
    \hat{a}\xrightarrow{p} a _0
    &\text{ and }
    \sqrt{n}(\hat{a} - a _0) \xrightarrow{d} N(0,(G_{11}'\Sigma^{-1}G_{11})^{-1}), \\
    \hat{R}\xrightarrow{p} R _0
    &\text{ and }
    \sqrt{n}(\hat{R} - R _0) \xrightarrow{d} N(0,\Lambda + \Lambda \Gamma \Sigma \Gamma'\Lambda).
\end{aligned}
\end{equation}
\end{theorem}
In particular, $\Sigma$ can be calculated analytically:
\begin{equation}\label{Sigma}
    \Sigma = \mathrm{E}[hh'] = 
    [\Phi_{kl}(\rho_{i_2j_2}^{xx})-p_{i_2,k}p_{j_2,l}]_{i_2,j_2,k,l}.
\end{equation}
Again, the estimated values for variance can be obtained by substituting $\hat{a},\hat{R}$. The well-posedness condition of the equations is similar to the above, and the derivation of the asymptotic distribution only uses Taylor expansion. Similar proofs can be seen in Pierce (1982), Lee and Poon (1987).   

A simple example for this chapter is shown in Appendix2\ref{app2}. 

\section{Simulations}
In this section, we perform numerical simulations on the two-step method, that is, comparing Algorithm\ref{alg2}2 with the maximum likelihood method. The simulation consists of two parts, exploring the accuracy and speed of the algorithm respectively. 

Since there is no convenient function directly using MLE to estimate the correlation matrix and its covariance matrix, we use the \textsf{sem} function (structural equation model) in the \textbf{lavaan} package (Rosseel (2012)) in R for indirect estimation.  

For each step GMM optimization, we choose BFGS method and implement it with the \textsf{optim} function in R. All simulations are carried out on a computer with an Intel (R) Core (TM) i7-10700 CPU @ 2.90GH and 64G memory.

In the first simulation, we set up two $Y$s and $X$s similar to Appendix\ref{app2}2. The true distribution is:
\begin{equation*}
\begin{pmatrix}
    Y_1\\
    Y_2\\
    X_1\\
    X_2
\end{pmatrix}
\Leftarrow
\begin{pmatrix}
    Y_1\\
    Y_2\\
    Z_1\\
    Z_2
\end{pmatrix}
\sim
N(0,
\left[
 \begin{array}{cc:cc}
 1&&&\\
 0.3 &1&&\\ \hdashline
 0.4 &0.6 &1&\\
 0.5 &0.7 &0.8 &1
 \end{array}
 \right])
\end{equation*}
with $X_1,X_2$ both binary, and the thresholds are $a=b=0$. We set sample size $n=100,500,1000$ and the number of Monte Carlo $N=10000$, to calculate the sample mean and covariance of $\hat{R}_i$ series, the sample mean of estimators of covariance matrix.  The simulation results are shown in Table1\ref{table1}.    

\begin{table}\label{table1}
  \centering
  \renewcommand{\arraystretch}{0.6}
  \begin{tabular}{ccccccccccccc}
    \toprule
    \multicolumn{1}{c}{} &\multicolumn{6}{c}{Two-Step IGMM} &\multicolumn{6}{c}{MLE in \textbf{lavaan}}\cr
    \cmidrule(lr){2-7} \cmidrule(lr){8-13}
     &$\hat{\rho}_{12}^{yy}$ &$\hat{\rho}_{11}^{yx}$ &$\hat{\rho}_{12}^{yx}$ &$\hat{\rho}_{21}^{yx}$ &$\hat{\rho}_{22}^{yx}$ &$\hat{\rho}_{12}^{xx}$ &$\hat{\rho}_{12}^{yy}$ &$\hat{\rho}_{11}^{yx}$ &$\hat{\rho}_{12}^{yx}$ &$\hat{\rho}_{21}^{yx}$ &$\hat{\rho}_{22}^{yx}$ &$\hat{\rho}_{12}^{xx}$\\
    \midrule
    \multicolumn{1}{c}{} &\multicolumn{12}{c}{n=100, MEAN$\times$1e-4, COVR MCOV$\times$1e-5}\\
    \cmidrule(lr){2-13}
    \multicolumn{1}{c}{} &\multicolumn{6}{c}{time = 115.46s}  &\multicolumn{6}{c}{time = 554.06s}\\
    MEAN &2917&3987&4978&5965&6965&7972 &2967 &3995 &4982 &5977&6965&7973\\
    \cmidrule(lr){2-13}
    \multirow{6}{*}{COVR} &1146 &&&&&&1079\\
    &0656 &1474 &&&&&0625&1408\\
    &0715 &0919 &1406 &&&&0686&0871&1342\\
    &0467 &0257 &0184 &1291 &&&0438&0230&0166&1204\\
    &0494 &0193 &0133 &0828 &1165 &&0462&0173&0116&0750&1060\\
    &0131 &0271 &0126 &0351 &0194 &0628 &0123&0258&0115&0327&0176&0598\\
    \cmidrule(lr){2-13}
    \multirow{6}{*}{MCOV} &1008 &&&&&&1103\\
    &0589 &1339 &&&&&0625&1420\\
    &0635 &0816 &1254 &&&&0681&0871&1355\\
    &0410 &0219 &0150 &1149 &&&0457&0220&0155&1257\\
    &0436 &0169 &0116 &0741 &1026 &&0488&0172&0120&0805&1125\\
    &0103 &0239 &0096 &0314 &0170 &0591 &0109&0241&0098&0320&0171&0608\\
    \midrule
    \multicolumn{1}{c}{} &\multicolumn{12}{c}{n=500, MEAN$\times$1e-4, COVR MCOV$\times$1e-6}\\
    \cmidrule(lr){2-13}
    \multicolumn{1}{c}{} &\multicolumn{6}{c}{time = 101.45s}  &\multicolumn{6}{c}{time = 618.37s}\\
    MEAN &2986&3996&4996&5988&6988&7995 &2990&4005&4998&5996&6991&7996\\
    \cmidrule(lr){2-13}
    \multirow{6}{*}{COVR} &2191 &&&&& &2172\\
    &1233 &2780 &&&&&1228&2743\\
    &1358 &1693 &2652 &&&&1355&1665&2616\\
    &0863 &0431 &0290 &2471 &&&0861&0424&0283&2393\\
    &0929 &0357 &0255 &1589 &2207 & &0927&0349&0250&1503&2107\\
    &0207 &0491 &0197 &0651 &0363 &1185 &0201&0480&0188&0620&0335&1161\\
    \cmidrule(lr){2-13}
    \multirow{6}{*}{MCOV} &2151 &&&&&&2183\\
    &1237 &2796 &&&&&1248&2821\\
    &1337 &1685 &2618 &&&&1352&1691&2642\\
    &0864 &0459 &0314 &2398 &&&0879&0453&0312 &2403\\
    &0921 &0353 &0242 &1536 &2141 & &0939 &0346  &0239 &1506 &2112\\
    &0214 &0494 &0197 &0647 &0350 &1163 &0216&0494&0196&0635&0335&1172\\
    \midrule
    \multicolumn{1}{c}{} &\multicolumn{12}{c}{n=1000, MEAN$\times$1e-4, COVR MCOV$\times$1e-6}\\
    \cmidrule(lr){2-13}
    \multicolumn{1}{c}{} &\multicolumn{6}{c}{time = 114.84s}  &\multicolumn{6}{c}{time = 686.64s}\\
    MEAN &2992&4002&4998&5991&6988&8001&2997&4003&5000&5993&6989&7997\\
    \cmidrule(lr){2-13}
    \multirow{6}{*}{COVR} &1092 &&&&&&1088\\
    &0619 &1399 &&&&&0618&1397\\
    &0666 &0833 &1314 &&&&0663&0820&1297\\
    &0434 &0240 &0157 &1225 &&&0433&0236&0154&1192\\
    &0468 &0192 &0125 &0790 &1097 & &0468&0191&0126&0752&1055\\
    &0098 &0255 &0104 &0318 &0170 &0582 &0097&0251&0101&0303&0156&0571\\
    \cmidrule(lr){2-13}
    \multirow{6}{*}{MCOV} &1082 &&&&&&1090\\
    &0621 &1404 &&&&&0623&1408\\
    &0671 &0845 &1315 &&&&0675&0842&1315\\
    &0434 &0230 &0157 &1204 &&&0437&0226&0156&1194\\
    &0462 &0177 &0121 &0771 &1076 & &0467&0173&0119&0746&1047\\
    &0107 &0248 &0099 &0324 &0175 &0579 &0108&0247&0098&0316&0167&0582\\
    \bottomrule
  \end{tabular}
  \caption{Comparison of MLE and Algorithm2\ref{alg2}, with MEAN the mean, COVR the covariance matrix of Monte Carlo $\hat{R}$ series, MCOV the mean of estimator series of Var$\hat{R}$.}
\end{table}

From Table1\ref{table1} we can conclude that:
\begin{itemize}
    \item The two-step IGMM estimator shows consistency ($\hat{R}$ are unbiased and variance decreases as sample size $n$ increases); 
    \item The IGMM is much faster than MLE, and the increase in $n$ will not significantly increase its time-consuming, since we only used some simple moments to build the equations;
    \item The variance of IGMM is slightly larger than MLE, but the gap decreases as $n$ increases. Theoretically, GMM is asymptotically efficient, and its efficiency is slightly lower than MLE;
    \item The variance of IGMM will be slightly underestimated than the true variance, and the gap decreases as $n$ increases, since Theorem\ref{the2} describes the large sample property.
\end{itemize}

To study performance of the algorithm2\ref{alg2} when there are more polychoric coefficients, we set the true distribution of the second simulation as:
\begin{equation*}
\begin{pmatrix}
    Y_1\\
    Y_2\\
    X_1\\
    X_2\\
    X_3
\end{pmatrix}
\Leftarrow
\begin{pmatrix}
    Y_1\\
    Y_2\\
    Z_1\\
    Z_2\\
    Z_3
\end{pmatrix}
\sim
N(0,
\left[
 \begin{array}{cc:ccc}
 1&&&\\
 -0.4 &1&&\\ \hdashline
 -0.3 &0.0 &1&\\
 -0.2 &0.1 &0.3 &1\\
 -0.1 &0.2 &0.4 &0.5 &1
 \end{array}
 \right]),
\end{equation*}
with $X_1,X_2,X_3$ all ternary. The thresholds are $-0.431,0.431$. The sample size $n=1000$ and Monte Carlo $N=10000$. The results are shown in Table2\ref{table2}:
\begin{table}\label{table2}
  \centering
  \renewcommand{\arraystretch}{0.6}
  \begin{tabular}{ccccccccccccc}
    \toprule
     &$\hat{\rho}_{12}^{yy}$ &$\hat{\rho}_{11}^{yx}$ &$\hat{\rho}_{12}^{yx}$ &$\hat{\rho}_{13}^{yx}$ &$\hat{\rho}_{21}^{yx}$ &$\hat{\rho}_{22}^{yx}$ &$\hat{\rho}_{23}^{yx}$ &$\hat{\rho}_{12}^{xx}$ &$\hat{\rho}_{13}^{xx}$ &$\hat{\rho}_{23}^{xx}$ \\
    \midrule
    \multicolumn{1}{c}{} &\multicolumn{10}{c}{Two-step IGMM, time=288.24s}\\
    \cmidrule(lr){2-13}
    MEAN &-3977&-3003&-2012&-1018&0011&1006&2001&3026&4034&5028\\
    \cmidrule(lr){2-13}
    \multirow{10}{*}{COVR} &1180 \\
    &0145 &1251 \\
    &0198 &0332 &1267 \\
    &0252 &0430 &0520 &1288 \\
    &-0328 &-0535 &-0152 &-0249 &1284\\
    &-0254 &-0113 &-0505 &-0237 &0306 &1307\\
    &-0177 &-0157 &-0187 &-0494 &0394 &0523 &1225\\
    &-0035 &-0180 &-0300 &-0148 &0134 &-0032 &-0004 &1412\\
    &-0071 &-0051 &-0106 &-0338 &0255 &0001 &-0024 &0507 &1257\\
    &-0053 &-0061 &-0064 &-0172 &0066 &0201 &0045 &0323 &0165 &1036\\
    \cmidrule(lr){2-13}
    \multirow{10}{*}{MCOV} &1120 \\
    &0119 &1156 \\
    &0173 &0319 &1196 \\
    &0233 &0389 &0500 &1218 \\
    &-0294 &-0492 &-0144 &-0214 &1226\\
    &-0234 &-0115 &-0476 &-0224 &0295 &1218\\
    &-0174 &-0159 &-0198 &-0477 &0383 &0502 &1196\\
    &-0022 &-0161 &-0298 &-0150 &0113 &-0025 &-0008 &1319\\
    &-0050 &-0021 &-0104 &-0292 &0214 &0007 &-0057 &0443 &1149\\
    &-0027 &-0038 &-0032 &-0162 &0039 &0162 &0032 &0284 &0142 &0944\\
    \midrule
    \multicolumn{1}{c}{} &\multicolumn{10}{c}{MLE in \textbf{lavaan}, time=825.29s}\\
    \cmidrule(lr){2-13}
    MEAN &-4004&-3001&-2008&-1007&0007&1003&2001&3001&4008&4997\\
    \cmidrule(lr){2-13}
    \multirow{10}{*}{COVR} &1149 \\
    &0137 &1216 \\
    &0194 &0319 &1238 \\
    &0245 &0419 &0507 &1255 \\
    &-0320 &-0518 &-0146 &-0242 &1255\\
    &-0246 &-0109 &-0491 &-0234 &0300 &1269\\
    &-0172 &-0156 &-0182 &-0484 &0387 &0507 &1195\\
    &-0034 &-0177 &-0293 &-0140 &0131 &-0029 &-0005 &1376\\
    &-0070 &-0052 &-0106 &-0327 &0249 &-0001 &-0024 &0492 &1221\\
    &-0051 &-0060 &-0063 &-0171 &0063 &0192 &0046 &0316 &0160 &1011\\
    \cmidrule(lr){2-13}
    \multirow{10}{*}{MCOV} &1165 \\
    &0125 &1192 \\
    &0178 &0327 &1228 \\
    &0238 &0399 &0509 &1249 \\
    &-0299 &-0499 &-0145 &-0217 &1256 \\
    &-0239 &-0116 &-0486 &-0227 &0299 &1248\\
    &-0179 &-0159 &-0199 &-0486 &0389 &0511 &1227\\
    &-0024 &-0174 &-0311 &-0161 &0118 &-0020 &-0003 &1364\\
    &-0052 &-0033 &-0112 &-0307 &0222 &0013 &-0051 &0468 &1198\\
    &-0030 &-0044 &-0042 &-0175 &0047 &0175 &0042 &0310 &0166 &1000\\
    \bottomrule
  \end{tabular}
  \caption{Comparison of MLE and Algorithm2\ref{alg2}, with MEAN$\times$1e-4 the mean, COVR$\times$1e-6 the covariance matrix of Monte Carlo $\hat{R}$ series, MCOV$\times$1e-6 the mean of estimator series of Var$\hat{R}$. Sample size $n=1000$.}
\end{table}

From Table2\ref{table2} we observe that:
\begin{itemize}
    \item The estimators still show consistency;
    \item The speed of IGMM maintains the advantage over MLE;
    \item There is a slight under-estimation of the variance estimation of IGMM. When there are more parameters, more samples are needed to make this underestimation negligible.
\end{itemize}
Through the simulations in this section, we can conclude that two-step IGMM is consistent in estimating the mixed correlation coefficient, and it is much faster than MLE.

\section{Data Analysis}
In this section, we apply previous method to a real dataset which came from the study of investigation on parenteral nutrition in hospitalized patients with digestive tract tumors. The study was initiated by the Committee of Cancer Rehabilitation and Palliative Care of China in 2017. The main purpose of the investigation was to obtain nutritional risks and malnutrition prevalence among newly admitted cancer inpatients through nutritional risk screening and assessment. We select the following five variables as examples for calculating the mixed correlation coefficient matrix:
\begin{itemize}
    \item PGSGA. Patient-Generated Subjective Global Assessment (Jager and Ottery (2017)), continuous variable;
    \item Age. Continuous variable;
    \item BMI. Body Mass Index, continuous variable;
    \item PN. Parenteral Nutrition. An indicator indicating parenteral or enteral nutrition (1 for parenteral, 0 for enteral);
    \item NRS. Nutritional Risk Screening (Reber, etc. (2019)), ordinal variable (0-6).
\end{itemize}
The number of subjects is $1086$ and the sample size is $n=981$ after excluding outliers and missing data. Our main purpose is to explore the correlation between covariates and PGSGA. We scaled the continuous variables, and the estimation results by Two-Step IGMM are shown in Table3:
\begin{table}\label{table3}
  \centering
  \renewcommand{\arraystretch}{0.6}
  \begin{tabular}{ccccccccccccc}
    \toprule
     &$\hat{\rho}_{12}^{yy}$ &$\hat{\rho}_{13}^{yy}$ &$\hat{\rho}_{23}^{yy}$ &$\hat{\rho}_{11}^{yx}$ &$\hat{\rho}_{12}^{yx}$ &$\hat{\rho}_{21}^{yx}$ &$\hat{\rho}_{22}^{yx}$ &$\hat{\rho}_{31}^{yx}$ &$\hat{\rho}_{32}^{yx}$ &$\hat{\rho}_{12}^{xx}$ \\
    \cmidrule(lr){2-13}
    EST &0673&-3623&-0913&-0177&2694&-0701&1122&1198&-2405&1123\\
    \cmidrule(lr){2-13}
    \multirow{10}{*}{COV} &0951 \\
    &0008 &0773 \\
    &-0393 &0065 &0954 \\
    &-0147 &0134 &0052 &1594 \\
    &0072 &-0165 &0002 &0103 &0792\\
    &-0157 &0009 &0198 &0196 &0062 &1794\\
    &0200 &0026 &-0188 &-0002 &0109 &0095 &0956\\
    &0002 &0017 &-0061 &-0593 &-0066 &-0192 &-0038 &1291\\
    &0000 &0239 &0117 &-0008 &-0322 &-0040 &-0134 &0068 &0837\\
    &-0002 &0043 &0086 &0381 &0065 &0349 &-0181 &-0401 &0207  &1706\\
    \bottomrule
  \end{tabular}
  \caption{Two-Step IGMM estimation results of the mixed correlation coefficient matrix of Parenteral Nutrition data, with EST$\times$1e-4 the estimated value, COV$\times$1e-6 the covariance matrix. Sample size $n=981$.}
\end{table}

We also use the scatterplot of matrices (SPLOM) to visualize the correlation matrix, with bivariate scatterplots below the diagonal, histograms on the diagonal, and the mixed correlation coefficients with standard errors above the diagonal. Correlation ellipses are drawn in the same graph. The red lines below the diagonal are the LOESS smoothed lines, fitting a smooth curve between two variables. As shown in Figure 1\ref{MLEIGMM}, the left panel is the result of Two-Step IGMM while the right is that of \textbf{lavvan} in R:

**************

Figure 1 Here

**************

Comparing the two figures, the IGMM method in this article is very close to the target MLE results in both numerical value and significance. Parenteral nutrition has no significant correlation with the response variable PGSGA, but it has a significant positive impact on BMI and NRS, which have high correlation with PGSGA. Therefore, it is speculated that BMI and NRS may be mediate or confounder variables between PN and PGSGA, thus a SEM with PN$\rightarrow$BMI/NRS$\rightarrow$PGSGA causal path can be established for further inference.

\section{Conclusion and Discussion}
In this article, we propose a new method for calculating the mixed correlation coefficient matrix and its covariance matrix based on the GMM framework. 

When modeling latent variables with SEM, if the observed data include both continuous and ordinal variables, it is necessary to estimate the Pearson, polyserial and polychoric coefficients simultaneously and calculate their asymptotic distribution. The popular MLE is slow due to high-dimensional integration in likelihood function. 

We build moment equations for each coefficient and align them into a GMM system. The integral is set to double integrals and approximated with Legendre polynomials. Finally, a modular and faster Two-Step IGMM is constructed. Theory and simulation show that Two-Step IGMM has consistency and asymptotic normality, and its efficiency is asymptotically equivalent to MLE. On this basis, it has faster calculation speed and more flexible model settings (Since the moment equations block for each coefficient is separate, you can only include the coefficients of interest instead of the entire correlation matrix). Moreover, when full data is not available, you can complete the entire estimation based on only some cross moments, which have good application prospects in meta-analysis.

The Pearson and polyserial correlation coefficients are both simple linear regression in GMM equations system, but polychoric is a nonlinear equation. We wonder if all three types of coefficients can be written as linear regression so that the entire mixing correlation matrix can be calculated directly using IRLS, and avoid all multiple integral. This is our future research direction.

\newpage
\centerline{Figure Environments}

\medskip

\begin{manfigure}\label{MLEIGMM}
\centering
\subfigure[Two-Step IGMM]{
    \fbox{\includegraphics[width=0.45\textwidth]{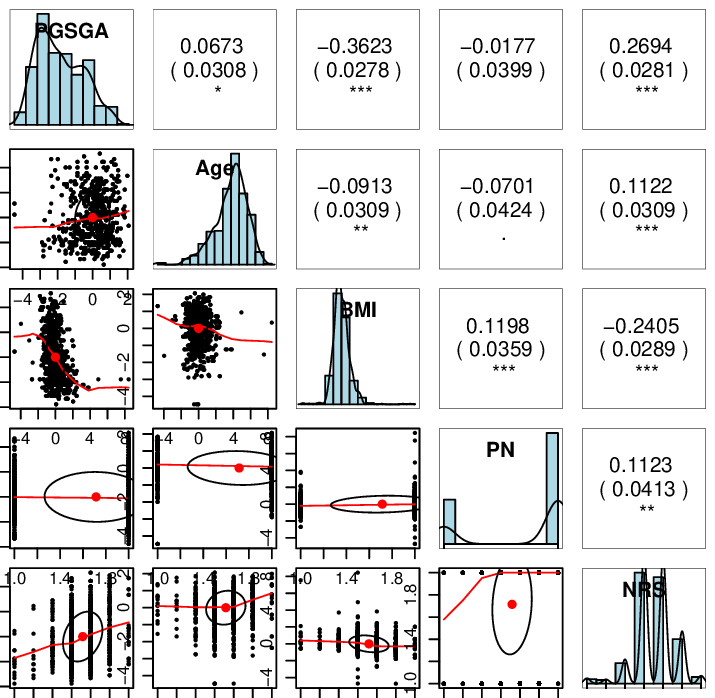}}
}
\subfigure[MLE in \textbf{lavvan}]{
    \fbox{\includegraphics[width=0.45\textwidth]{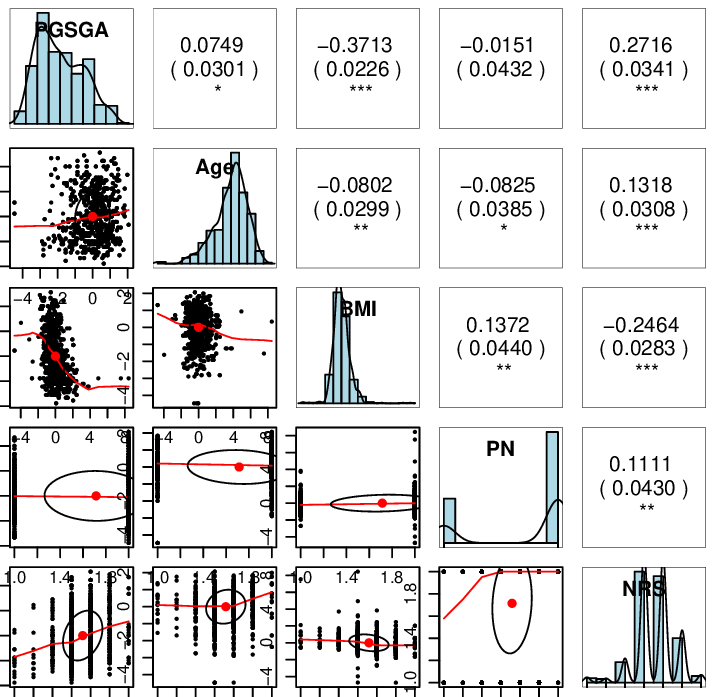}}
}
\caption{The SPLOM of Two-Step IGMM and MLE with the parenteral nutrition data. * indicate significance level.}
\end{manfigure}

\vspace{\fill}

\newpage
\appendix

\section{Gradient Matrix}\label{app1}
The moment equations $u(\cdot)$ are aligned according to parameters, so most of the gradient matrix is block-diagonal. For $G_{11}$, each block is a bidiagonal matrix, and its lower subdiagonal is the negative of the upper element:
\begin{equation*}
    G_{11} = 
    -\text{diag}_{i_2}(
\begin{bmatrix}
    \phi(a_{i_2,1}) \\
    -\uparrow &\phi(a_{i_2,2})\\
    \vdots &\ddots &\ddots\\
    0 &\ldots &-\uparrow &\phi(a_{i_2,s_{i_2}-1})
\end{bmatrix}).
\end{equation*}
Obviously $G_{12}=\mathbf{0}$. Since $g(\cdot)$ is segmented into three parts, the derivative of $g(\cdot)$ can also be calculated separately. For Pearson coefficient:
\begin{equation*}
    G_{21}^{yy} = \mathbf{0}, G_{22}^{yy} = -\mathbf{I},
\end{equation*}
with $\mathbf{I}$ corresponding scale identity matrix. 

For polyserial, fixing the $i_1, i_2$, the matrix is similar with $G_{11}$'s form except for one more row. Then block diagonalize it according to $i_2$ and bind it in row according to $i_1$:
\begin{equation*}
G_{21}^{yx} = -\text{rbind}_{i_1}(\text{diag}_{i_2}(\rho_{i_1 i_2}^{yx}
\begin{bmatrix}
    a_{i_2,1}\phi(a_{i_2,1}) \\
    -\uparrow &a_{i_2,2}\phi(a_{i_2,2})\\
    \vdots &\ddots &\ddots\\
    0 &\ldots &-\uparrow &a_{i_2,s_{i_2}-1}\phi(a_{i_2,s_{i_2}-1})\\
    0 &\ldots &0 &-\uparrow
\end{bmatrix})).
\end{equation*}
For $G_{22}$, firstly arrange $\xi$'s into column vector, and then diagonalize it according to $i_2$ and $i_1$ successively:
\begin{equation*}
G_{22}^{yx} = -\text{diag}_{i_1}(\text{diag}_{i_2}(\text{vector}_k(\xi_{i_2,k}))).
\end{equation*}

In polychoric parts, the derivative with respect to threshold is:
\begin{equation*}
    \frac{\partial\Phi_{kl}(\rho_{i_2 j_2}^{xx})}{\partial a_{i_2,k}} = \phi(a_{i_2,k})[\Phi(\frac{a_{j_2,l}-\rho a_{i_2,k}}{\sqrt{1-\rho^2}}) - \Phi(\frac{a_{j_2,l-1}-\rho a_{i_2,k}}{\sqrt{1-\rho^2}})]:= \zeta_{kl}(a_{i_2,k},\rho_{i_2 j_2}^{xx}),
\end{equation*}
with $\zeta(,)$ represent the function above. For a certain equation belong to $\rho_{i_2 j_2}^{xx}$ in cell $k,l$, the derivative of four thresholds can be denote by $\zeta(,)$ while upper thresholds positive and lower thresholds negative. Then:
\begin{equation*}
    G_{21}^{xx} = \text{rbind}_{i_2,j_2}(
    \begin{bmatrix}
        -\zeta_{kl}(a_{i_2,k-1},\rho_{i_2 j_2}^{xx}) &
        \zeta_{kl}(a_{i_2,k},\rho_{i_2 j_2}^{xx})
    \end{bmatrix}
    \begin{bmatrix}
        -\zeta_{kl}(a_{i_2,l-1},\rho_{i_2 j_2}^{xx}) &
        \zeta_{kl}(a_{i_2,l},\rho_{i_2 j_2}^{xx})
    \end{bmatrix}
    ).
\end{equation*}
That's, there is a bidiagonal matrix similar to $G_{21}^{yx}$ structure at the block corresponding to $X_{i_2},X_{j_2}$, then bind the blocks by row. Finally,
\begin{equation*}
    \Phi_{kl}(\rho_{i_2 j_2}^{xx}) = \Phi(a_{i_2,k},a_{j_2,l})-\Phi(a_{i_2,k},a_{j_2,l-1})-\Phi(a_{i_2,k-1},a_{j_2,l}) + \Phi(a_{i_2,k-1},a_{j_2,l-1}),
\end{equation*}
with $\Phi(,)$ bivariate normal distribution function with correlation $\rho_{i_2 j_2}^{xx}$. And
\begin{equation*}
    \frac{\partial\Phi_{kl}(\rho_{i_2 j_2}^{xx})}{\partial\rho_{i_2 j_2}^{xx}} = \phi(a_{i_2,k},a_{j_2,l})-\phi(,)-\phi(,) + \phi(,):=\phi_{kl}(\rho_{i_2 j_2}^{xx}),
\end{equation*}
with $\phi(,)$ bivariate normal density function with correlation $\rho_{i_2 j_2}^{xx}$. Then,
\begin{equation*}
    G_{22}^{xx} = -\text{diag}_{i_2}(
    \text{vector}_{j_2}(\phi_{kl}(\rho_{i_2 j_2}^{xx})
    ).
\end{equation*}

Combining all blocks, we have:
\begin{equation*}
    G = 
    \left[
    \begin{array}{c:ccc}
        G_{11} \\ \hdashline
         & -\mathbf{I}\\
        G_{21}^{yx} & &G_{22}^{yx}\\
        G_{21}^{xx} & &&G_{22}^{xx}
    \end{array}\right]. 
\end{equation*}
$G$ must be consistent in dimension with the equation system\ref{totalmoment}, thus corresponding rows should be removed. The matrix formula is complicated, but through the following examples\ref{app2} you can find that each block is very clear.

\newpage

\section{A Simplest Example}\label{app2}
A minimum set containing Pearson, polyserial and polychoric correlation coefficients is the four variables system $Y_1,Y_2,X_1,X_2$, 
\begin{equation*}
\begin{pmatrix}
    Y_1\\
    Y_2\\
    X_1\\
    X_2
\end{pmatrix}
\Leftarrow
\begin{pmatrix}
    Y_1\\
    Y_2\\
    Z_1\\
    Z_2
\end{pmatrix}
\sim
N(0,
 \begin{bmatrix}
 1\\
 \rho^{yy}_{12} &1\\
 \rho^{yx}_{11} &\rho^{yx}_{21} &1\\
 \rho^{yx}_{12} &\rho^{yx}_{22} &\rho^{xx}_{12} &1
 \end{bmatrix})\\
\end{equation*}
with $X_1,X_2$ both binary, and the thresholds are $a$ and $b$ respectively, that's:
\begin{equation*}
\begin{aligned}
    X_1 &= 1 \text{ if } Z_1\leq a\\
    X_1 &= 2 \text{ if } Z_1 > a\\
    X_2 &= 1 \text{ if } Z_2\leq b\\
    X_2 &= 2 \text{ if } Z_2> b,
\end{aligned}
\end{equation*}
and $\rho_{12}^{xx}$ is a tetrachoric correlation. Thus the parameters to estimated is $(a,b)',R=(\rho^{yy}_{12},\rho^{yx}_{11},\rho^{yx}_{12},\rho^{yx}_{21},\rho^{yx}_{22},\rho^{xx}_{12})'$. The moment equations for thresholds are:
\begin{equation*}
\mathrm{E}[h(\cdot)] = \mathrm{E}
\begin{bmatrix}
    \mathrm{I}_1(X_1)-\Phi(a)  \\
     \mathrm{I}_1(X_2)-\Phi(b) 
\end{bmatrix}=0.
\end{equation*}
The equations for correlations are:
\begin{equation*}
\mathrm{E}[g(\cdot)]=\mathrm{E}
\left[
\begin{array}{c}
     Y_1Y_2 - \rho_{12}^{yy}\\ \hdashline
     Y_1\mathrm{I}_1(X_1) + \rho_{11}^{yx}\phi(a)\\ 
     Y_1\mathrm{I}_2(X_1) - \rho_{11}^{yx}\phi(a)\\
     Y_1\mathrm{I}_1(X_2) + \rho_{12}^{yx}\phi(b)\\ \hline
     Y_2\mathrm{I}_1(X_1) + \rho_{21}^{yx}\phi(a)\\
     Y_2\mathrm{I}_2(X_1) - \rho_{21}^{yx}\phi(a)\\
     Y_2\mathrm{I}_1(X_2) + \rho_{22}^{yx}\phi(b)\\ \hdashline
     \mathrm{I}_{11}(X_1X_2) - \Phi(a,b,\rho_{12}^{xx})\\
     \mathrm{I}_{12}(X_1X_2) - \Phi(a)+\Phi(a,b,\rho_{12}^{xx})\\
     \mathrm{I}_{21}(X_1X_2) - \Phi(b)+\Phi(a,b,\rho_{12}^{xx})
\end{array}
\right]
=0.
\end{equation*}
The dash lines segment the three correlation equations. The solid line cuts the polyserial equations into $Y_1$ and $Y_2$ parts, which indicates that only a complete set of equations can be saved for the same $Y_i$. The gradients are: 
\begin{equation*}
    G_{11} = \frac{\partial h}{ \partial a}=
    \begin{bmatrix}
        -\phi(a)\\
        &-\phi(b)
    \end{bmatrix},
\end{equation*}
\begin{equation*}
    G_{21} = \frac{\partial g}{ \partial a}= 
    \begin{bmatrix}
        \mathbf{0}\\ \hdashline
        G_{21}^{yx}\\ \hdashline
        G_{21}^{xx}
    \end{bmatrix}=
    \begin{bmatrix}
        0 &0\\ \hdashline
        -\rho_{11}^{yx}a\phi(a) &0\\
        \rho_{11}^{yx}a\phi(a) &0\\
        0 &-\rho_{12}^{yx}b\phi(b)\\ 
        -\rho_{21}^{yx}a\phi(a) &0\\
        \rho_{21}^{yx}a\phi(a) &0\\
        0 &-\rho_{22}^{yx}b\phi(b)\\ \hdashline
        -\phi(a)\Phi(\frac{b-\rho_{12}^{xx}a}{\sqrt{1-\rho_{12}^{xx2}}}) &-\phi(b)\Phi(\frac{a-\rho_{12}^{xx}b}{\sqrt{1-\rho_{12}^{xx2}}})\\
        -\phi(a)(1-\Phi(\frac{b-\rho_{12}^{xx}a}{\sqrt{1-\rho_{12}^{xx2}}})) &\phi(b)\Phi(\frac{a-\rho_{12}^{xx}b}{\sqrt{1-\rho_{12}^{xx2}}})\\
        \phi(a)\Phi(\frac{b-\rho_{12}^{xx}a}{\sqrt{1-\rho_{12}^{xx2}}}) &-\phi(b)(1-\Phi(\frac{a-\rho_{12}^{xx}b}{\sqrt{1-\rho_{12}^{xx2}}}))
    \end{bmatrix},
\end{equation*}
\begin{equation*}
    G_{22} = \frac{\partial g}{ \partial R}= 
    \begin{bmatrix}
        -\mathbf{I}\\ 
        &G_{22}^{yx}\\ 
        &&G_{22}^{xx}
    \end{bmatrix}=
    \left[
    \begin{array}{c:cccc:c}
        -1 &&&&&\\ \hdashline
        &\phi(a) &&&& \\
        &-\phi(a) &&&&\\
        &&\phi(b) &&&\\
        &&&\phi(a) &&\\
        &&&-\phi(a) &&\\
        &&&&\phi(b)\\ \hdashline
        &&&&&-\phi(a,b,\rho_{12}^{xx})\\
        &&&&&\phi(a,b,\rho_{12}^{xx})\\
        &&&&&\phi(a,b,\rho_{12}^{xx})
    \end{array}\right].
\end{equation*}
The total gradient is combined as Appendix1\ref{app1}.

These are all components for one-step GMM. For two-step methods, an additional component (\ref{Sigma}) is:
\begin{equation*}
    \Sigma = \mathrm{E}[hh']=
    \begin{bmatrix}
        \Phi(a)(1-\Phi(a)) &\Phi(a,b,\rho_{12}^{xx})-\Phi(a)\Phi(b)\\
       \Phi(a,b,\rho_{12}^{xx})-\Phi(a)\Phi(b) &\Phi(b)(1-\Phi(b))
    \end{bmatrix}.
\end{equation*}

Then, the parameters estimation and inference can be performed in Algorithms1\ref{alg1} and Algorithms2\ref{alg2} based on the above components.

\end{document}